# Creating a Nanoscale Lateral Heterojunction in a Semiconductor Monolayer with a Large Built-in Potential

*Madisen Holbrook,[†] Yuxuan Chen,[†] Hyunsue Kim,[†] Lisa Frammolino,[†] Mengke Liu,[†] Chi-Ruei Pan,[‡] Mei-Yin Chou,[‡] Chengdong Zhang,[§] and Chih-Kang Shih[†,*]*

[†]Department of Physics, the University of Texas, Austin, Texas, USA 78712

[‡]Institute of Atomic and Molecular Sciences, Academia Sinica, Taipei 10617, Taiwan

[§]School of Physics and Technology, Wuhan University, Wuhan 430072, China

ABSTRACT: The ability to engineer atomically thin nanoscale lateral heterojunctions (HJs) is critical to lay the foundation for future two-dimensional (2D) device technology. However, the traditional approach to creating a heterojunction by direct growth of a heterostructure of two different materials constrains the available band offsets, and it is still unclear if large built-in potentials are attainable for 2D materials. The electronic properties of atomically thin semiconducting transition metal dichalcogenides (TMDs) are not static, and their exciton binding energy and quasiparticle band gap depend strongly on the proximal environment. Recent studies have shown that this effect can be harnessed to engineer the lateral band profile of monolayer

TMDs to create a heterojunction. Here we demonstrate the synthesis of a nanoscale lateral heterojunction in monolayer $MoSe_2$ by intercalating Se at the interface of a hBN/Ru(0001) substrate. The Se intercalation creates a spatially abrupt modulation of the local hBN/Ru work function, which is imprinted directly onto an overlying $MoSe_2$ monolayer to create a large built-in potential of $0.83 \pm 0.06$ eV. We spatially resolve the $MoSe_2$ band profile and work function using scanning tunneling spectroscopy to map out the nanoscale depletion region. The Se intercalation also modifies the dielectric environment, influencing the local band gap renormalization and increasing the $MoSe_2$ band gap by ~0.26 eV. This work illustrates that environmental proximity engineering provides a robust method to indirectly manipulate the band profile of 2D materials outside the limits of their intrinsic properties, providing avenues for future device design.

As technology has advanced, the rush to miniaturization has led to the ultimate limit of atomically thin, two dimensional materials. Within this family of materials, the emergence of 2D semiconducting TMDs has provided exciting possibilities for advanced 2D electronics, including devices comprised of a single atomic layer.[1] However, future technology based on TMDs hinges on the capability of creating nanoscale lateral junctions with large built-in potentials. Following the paradigm of three-dimensional counterparts, lateral heterojunctions have been directly synthesized by laterally stitching two different TMD monolayers together,[2-5] although the built-in potentials are on the order of hundreds of meV, limited by the TMD library.[6] Development of substitutional doping is challenging to spatially control while avoiding degradation due to the atomic thin nature of TMDs. However, TMDs are not limited to the toolbox of techniques conventionally used for bulk materials, as their extreme 2D nature provides other avenues to manipulate their electronic properties.  A variety of effects modify TMD electronic structure including (but not limited to) surface adsorbates,[7] substrate hybridization,[8-10] substrate work function,[11-13] and dielectric screening.[14-17]  Although this strong influence of the proximal environment creates design challenges, it can also be exploited as a tool to non-invasively engineer a lateral heterojunction in a monolayer TMD. Recently, TMD lateral HJs were "coulomb engineered" by creating an abrupt change in the dielectric environment to modify the local band gap renormalization, though the built-in potentials are relatively small, and the lateral length of the junction is unclear.[18-20] Similarly, surface adsorbates and ferroelectric substrates have been utilized to engineer TMD lateral HJs by electrostatically tuning the Fermi level position and work function.[7, 21, 22] We adopt the terminology of "heterojunction" rather than "homojunction" following the terminology of these prior studies; TMD monolayers are so influenced by the nearby environment, they can be considered different material systems on either side of the junction. The above examples illustrate that environmental proximity engineering is a promising direction for future TMD lateral HJ design and invite further exploration.

Here we create a nanoscale lateral heterojunction in a single $MoSe_2$ layer with a built-in potential of $0.83 \pm 0.06$ eV, utilizing a proximity engineering approach that centers on Se intercalation at the interface of a hBN/Ru(0001) heterostructure substrate. The interaction of hBN grown epitaxially on transition metal substrates such as Ru(0001) produces a so called corrugated "nanomesh" superstructure  that exhibits a periodic modulation of the work function.[23, 24]  It was later shown that this modulation translates directly to a corresponding periodic shift of ~0.15 eV of the band gap of a MoSe$_2$ monolayer grown epitaxially grown on top of the hBN/Ru(0001), signifying a purely electrostatic effect.[11]  Although hBN is normally a wide band gap insulator, the MoSe$_2$ on hBN/Ru(0001) also interestingly exhibits a smaller band gap than on graphite or graphene, suggesting a larger band gap renormalization due to the strong interaction between the hBN and Ru.[11] It is well established that intercalation of adsorbates alters the interaction between hBN on transition metal substrates by electronically decoupling the hBN layer, resulting in a large enhancement of the work function.[25-29] These prior studies provide the framework to develop a strategy to engineer a nanoscale lateral HJ with a large built-in potential by utilizing Se intercalation of a hBN/Ru platform to create an underlying electrostatic template that is imprinted on overlying MoSe$_2$. Here we not only show that intercalation can be used to engineer a lateral HJ, but we also examine the relationship of the work function modulation, substrate dielectric screening, and MoSe$_2$ band profile at the nanoscale.

### Results and Discussion

The hBN/Ru platform is synthesized using the well-documented UHV CVD technique (see Methods for details) and the resulting high quality hBN monolayer (see Figure 1a) shows the distinct corrugated superstructure as was previously reported.[24, 30]  Se intercalation is achieved by exposing the corrugated hBN/Ru to Se vapor in a molecular beam epitaxy system (see Methods), which changes the hBN surface morphology. As shown in Figure 1b, large flat regions emerge where the Se has intercalated into the hBN/Ru interface. Figure 1c allows closer inspection of an hBN intercalation boundary with a corresponding topographic profile that shows the apparent height of the hBN increases as the Se intercalant decouples the hBN from the Ru. The hBN overlayer obscures the intercalated Se on the Ru surface, therefore experimental determination of the intercalation phase is highly nontrivial. We performed first-principles simulation of Se adsorbed on the Ru(0001) surface and found that the HCP threefold hollow sites were the most energetically favorable adsorption site. Moreover, the most stable configuration is at a coverage of 1/9 HCP sites due to a repulsive mutual interaction between the Se atoms, similar to previous theoretical studies[31, 32] (see Supporting Information). Our calculations also show that upon Se intercalation, the corrugated hBN/Ru(0001) is transformed to an almost flat layer with the average interlayer distance increased by 1.9 Å (see Supporting Information), consistent with the topographic line cut from the STM image in Figure 1c. Our STM measurements further show that the Se intercalation alters the hBN work function, with an increase of ~0.85 eV, which we discuss below. We note that this change in the work function and surface morphology is a common phenomenon for intercalated hBN on other transition metal substrates, suggesting that this approach may be generalized.[25-28]

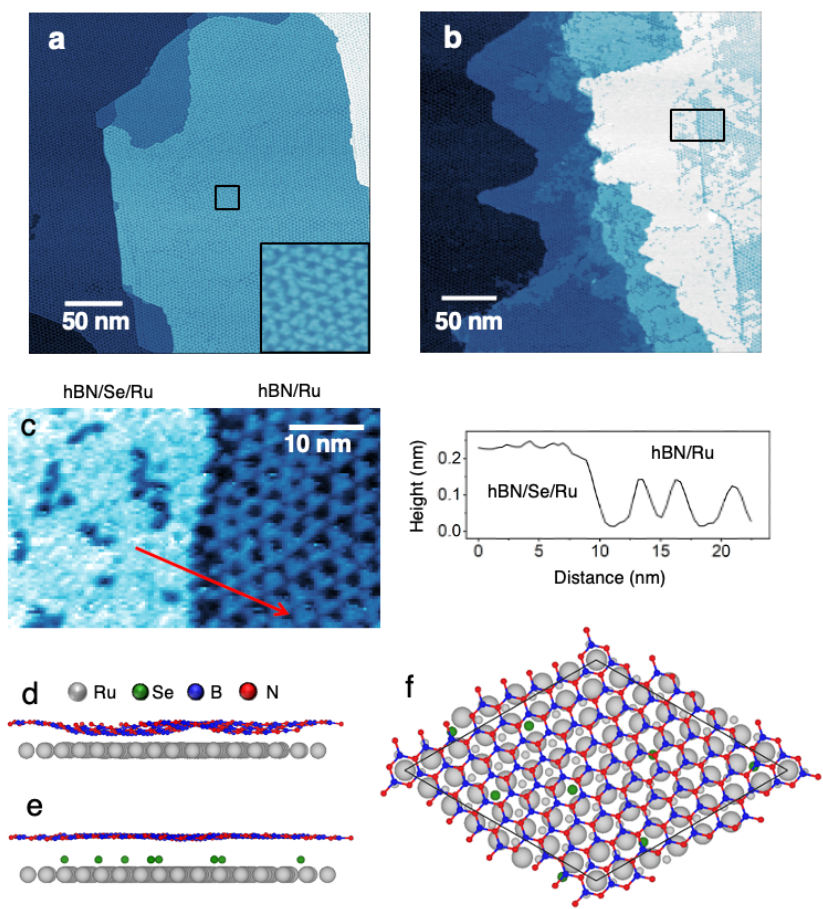

**Figure 1. (a)** Topographic STM image of hBN on Ru showing a corrugated structure (inset). **(b)** STM image of the hBN layer after Se intercalation, showing domains of a flat structure have emerged. **(c)** STM image of a Se intercalation heterointerface similar to the area indicated by the box in (b). To the right is a topographic line cut corresponding to the red vector. **(d, e)** Side view of simulated hBN/Ru(0001) and hBN/Se/Ru(0001) with a concentration of 11% Se (1/9 ML coverage of HCP sites) intercalation, respectively. **(f)** Top view of the simulated hBN/Se/Ru(0001) supercell. The STM images in (a,b,c) were obtained at a sample bias and tunneling current of V = -2 V and I = 10 pA.

The epitaxial $MoSe_2$ synthesis and Se intercalation is achieved with a direct one step method by applying a high Se overpressure during the $MoSe_2$ growth (see Methods). An STM topographic image of the heterostructure (Figure 2a) shows the coexistence of two structural phases, flat and corrugated, of both the $MoSe_2$ and the underlying hBN (see Supporting Information for an analysis of $MoSe_2$ growth). The growth of the hBN on Ru is self-limited to a monolayer (as shown by Fig. 1a),[11, 24] therefore islands on the hBN surface can be attributed to the $MoSe_2$ (shown in orange in Figure 2a). We note that the flat hBN/Se/Ru regions often lie next to the flat $MoSe_2$, showing a continuation of the Se intercalation. Thus, the overlying $MoSe_2$ mirrors the structural changes of the hBN, appearing flat in the Se intercalated regions, as shown by the atomic resolution STM images of the $MoSe_2$ surface (Figures 2b,c). From atomic resolution STM images, we measure a $MoSe_2$ lattice constant of $0.33 \pm 0.01$ nm, further confirming that the islands are indeed $MoSe_2$ (see Supporting Information). The Se intercalation is often concentrated to the Ru step edges, creating a long and straight intercalation boundary with a sharp interface (see Fig. 2a). The abrupt transition from the flat to corrugated morphology hints that the $MoSe_2$ electronic structure evolves over the same length scale, as discussed later.

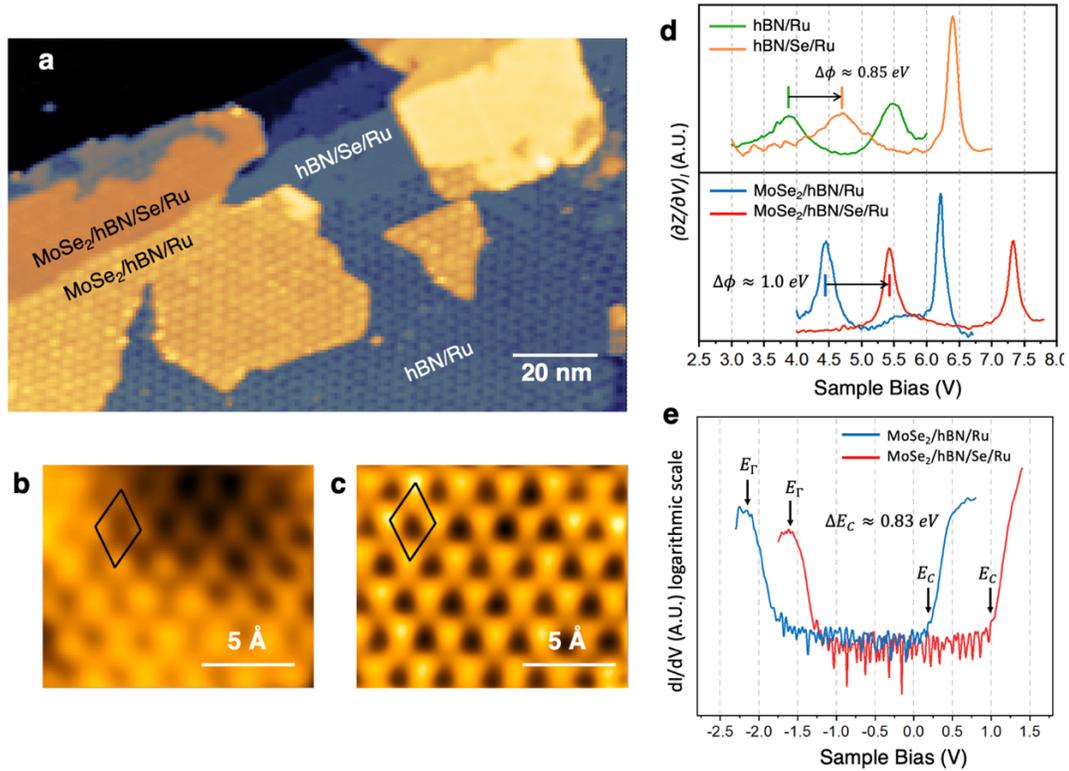

**Figure 2. (a)** STM image of MoSe₂ on Se intercalated hBN (flat) and unintercalated hBN (corrugated) on Ru. **(b,c)** Atomic resolution STM images of the corrugated MoSe₂ on hBN/Ru and flat MoSe₂ on hBN/Se/Ru, respectively. **(d)** dZ/dV FER spectroscopy on hBN/Ru (green) and hBN/Se/Ru (orange) (top) and on the MoSe₂/hBN/Ru (blue) and Se intercalated MoSe₂/hBN/Se/Ru (red) (bottom). We note that the measurements on the MoSe₂/hBN/Ru and hBN/Ru were taken on the "wire" regions of corrugated superstructure reported in Ref. 11. **(e)** dI/dV spectroscopy averaged from different locations on MoSe₂/hBN/Ru at "wire" regions (blue) and Se intercalated MoSe₂/hBN/Se/Ru (red). STM images obtained at a sample bias and current setpoint of (a) V = -2.6 V, I = 5 pA (b) V = -1.3 V, I = 0.25 nA (c) V = -1.3 V, I = 20 pA.

Next, we investigate how the Se intercalation of the hBN influences the electronic structure of the MoSe₂ overlayer. Shown in Figure 2d are $(\partial Z/\partial V)_I$ spectra acquired in the constant current mode revealing field emission resonances (FER) obtained on hBN/Ru and hBN/Se/Ru regions respectively. The onset of the first resonance of the FER spectrum is a good approximation for the sample work function[33, 34] (discussed in Supporting Information), therefore a change in the onset bias reflects a relative change in the work function. Figure 2d shows that upon Se-intercalation of the hBN/Ru, the work function is increased by 0.85 eV, recovering a value similar to that of clean Ru. A first-principles calculation of the work function for hBN/Ru with and without Se intercalation reveals that the Se intercalant increases the work function by 1.1 eV (see details in Supporting Information). The calculation also shows that the Se adsorption energy and hBN work function hardly change between 1/9 and 1/4 monolayer Se concentration, therefore we suggest that there is a range of intercalant concentrations that would yield the same result (see Supporting Information). Similar studies of intercalated hBN on transition metals have shown that the decoupling occurs at some critical intercalant coverage and further intercalant exposure does not alter the hBN electronic properties.[25, 27] This implies that the increase in work function is governed by the isolation of the hBN, which disrupts the strong charge transfer with the Ru, similar to previous reports for other intercalants of hBN/Ru.[26] Interestingly, FER measurements on the epitaxially grown $MoSe_2$ monolayers show a similar work function increase (Figure 2d bottom) due to the Se intercalation, signifying that the changes in electrostatic potential are mirrored in the $MoSe_2$. A difference of $\Delta\phi$ ($\sim 0.15$ eV) is likely due to the spatial fluctuation of the intercalation, as the FER of the $MoSe_2$ and that of the hBN are measured at different locations. Due to the modulation of the work function across the corrugated $MoSe_2$ and hBN (discussed above), it is important to note that the FER values reported here correspond to the reported "wire" regions of the corrugated nanomesh reported in Ref. 11.

We performed dI/dV tunneling spectroscopy to determine how the Se intercalation changes the band gap of the $MoSe_2$, as shown in Figure 2e. Averaged conductance spectra obtained on different locations on the intercalated $MoSe_2$ (red) and wire regions of the corrugated $MoSe_2$ (blue) show a dramatic shift of the $MoSe_2$ band gap. Following the method of Ugeda et. al. (see Ref. 14) we determined the CBM positions (black arrows) of the averaged spectra and find that the Se intercalation increases the CBM position by $0.83 \pm 0.06$ eV. This large upshift in energy is correlated with the increase in the substrate work function (discussed above), which tunes the $MoSe_2$ band profile via the electrostatic field effect. Taking into account that the VBM at the K point is 0.4 eV above the $\Gamma$ point,[14, 35] we determine a quasiparticle bandgap of ~1.94 eV at the corrugated region, which agrees with the previous reported value,[11] and ~2.2 eV at the Se intercalated region. This bandgap increase hints that the dielectric screening differs between the two regions, leading to a variation in the band gap renormalization. The above analysis underscores the power of Se intercalation as a method to tune the hBN/Ru electrostatic potential landscape, which strongly influences the electronic properties of the overlying $MoSe_2$ monolayer.

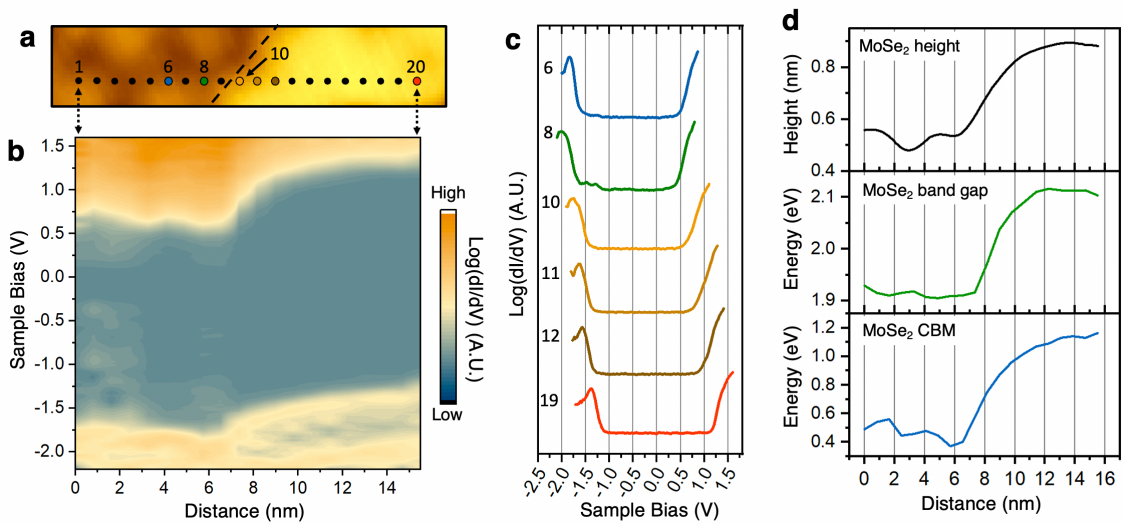

**Figure 3. (a)** STM image of an intercalation heterointerface between the $MoSe_2$/hBN/Ru and $MoSe_2$/hBN/Se/Ru obtained at a sample bias and current setpoint of V = -2 V and I = 10 pA. **(b)** Real space 2D color rendering of dI/dV spectroscopy obtained on the points from (a) along a 15.5 nm line **(c)** Select spectra from the line indicated in (a), showing the changes in the $MoSe_2$ bandgap across the intercalation boundary. **(d)** a height profile (top), plot of the $MoSe_2$ band gap (middle), and plot of the $MoSe_2$ CBM (bottom) extracted from the spectra in (b).

We next answer whether this large built-in potential is supported in the nanoscale regime by examining the detailed spatial evolution of the $MoSe_2$ band profile. We performed scanning tunneling spectroscopy (STS) dI/dV measurements along a line across the intercalation boundary shown in Figure 3a to spatially resolve the band gap. Figure 3b depicts a 2D color rendering of the real space mapping of the $MoSe_2$ band profile, exposing an unmistakable nanometer scale lateral heterojunction. This gives a qualitative view of the band bending behavior as the color scale does not reflect the band edge locations, which are clarified by the plot of select spectra from the line spectroscopy (Figure 3c). We note that the line cut slightly overlaps the corrugation, resulting in a modulation that shifts the bandgap (Figure 3c spectra 6 blue, 8 green) as was previously reported.[11] Close inspection of the spectra in Figure 3c unveils that the band gap renormalization begins abruptly at the interface (10, yellow) and is complete within a few nanometers (12, brown), while the shift in the band profile continues (19 red). Figure 3d shows a comparison of the $MoSe_2$ band gap, CBM position, and STM topographic profile along the line cut, elucidating the length scales of the band gap renormalization and the electrostatic shift of the band profile. This analysis shows that the renormalization of quasiparticle band gap is localized to the intercalation boundary and is complete within ~ 4 nanometers (middle panel), while the band bending continues. Recent theoretical investigations of coulomb engineered heterostructures predicted that band gap renormalization is completed within two unit cells,[36, 37] however, these calculations consider an atomically abrupt dielectric interface. Our theoretical calculations show that even an intercalation coverage of approximately 11% Se (1/9 of HCP sites), corresponding to an average Se-Se separation of ~1 nm, can lift the hBN layer from the Ru (See Supporting Information). This suggests that the nanometer scale of the $MoSe_2$ band gap renormalization reflects the length scale over which the hBN is decoupled from the Ru.

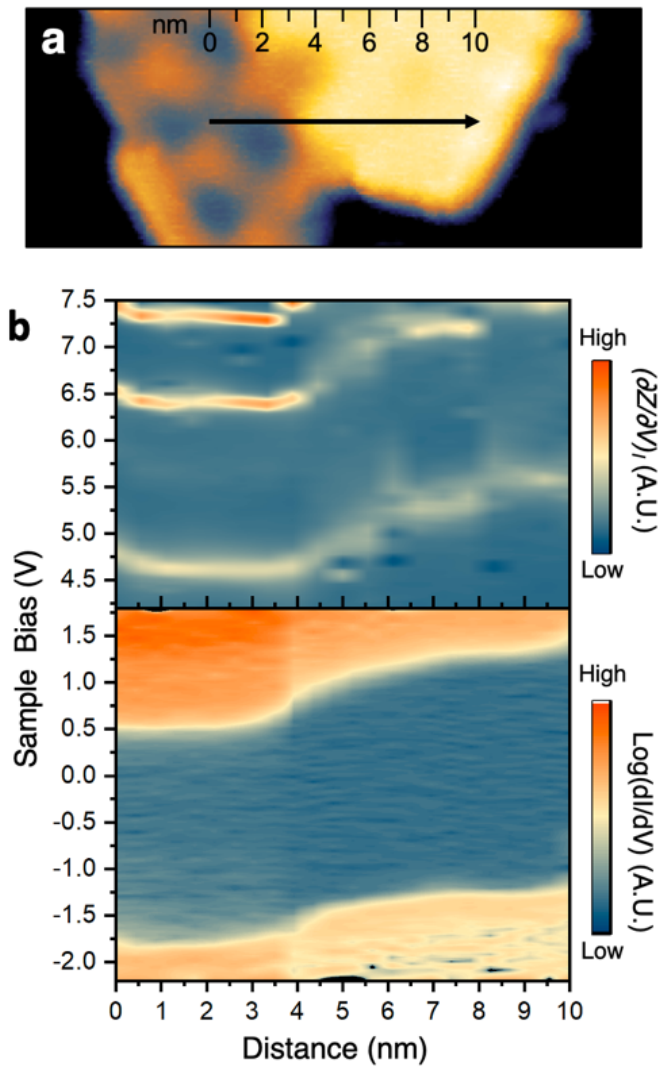

**Figure 4. (a)** STM image of a partially Se intercalated MoSe$_2$ island, sample bias V = -2 V, setpoint I = 10 pA. **(b)** 2D color rendering of $(\partial Z/\partial V)_I$ (top) and dI/dV (bottom) spectroscopy obtained along the black line in the STM image.

In order to characterize the spatial relationship between the local work function and the MoSe$_2$ band profile, we carry out FER $(\partial Z/\partial V)_I$ and STS dI/dV band mapping along the same line across the intercalation heterointerface (Figure 4a black vector). We note that the work function and band gap measurements discussed above (Figure 2) were obtained separately, thus we map the work function and band profile in tandem to address possible uncertainty due to spatial inhomogeneity of the intercalation. Figure 4b depicts a 2D color mapping of the spatially resolved FER $(\partial Z/\partial V)_I$ (top) and dI/dV spectroscopy (bottom), with a ~ 0.5 nm uncertainty in the alignment due to possible drift between measurements. Although there is some noise in the resulting FER measurement, the resulting energy profiles of the first FER peak $\phi(x)$ and the CBM $E_C(x)$ show a strikingly similar dispersion. This qualitatively shows that the MoSe$_2$ CBM tracks the work function, affirming that the MoSe$_2$ band bending is essentially an imprint of the changing electrostatic landscape. This suggests that the electron affinity $\chi = \phi - E_C$ is roughly constant (within experimental error), which is somewhat puzzling as band gap renormalization is also present. Note that the band bending of the VBM shows a different dispersion, which we suggest comes from the influence of the band gap renormalization. Theoretical investigations have predicted that band gap renormalization should be symmetric[18, 20, 36] and therefore the band gap change should be split evenly between the valence and conduction bands when referenced from the vacuum level. Our measurements suggest that this might not be the case, as symmetric band gap renormalization should result in a variation in $\chi$. However, we note that better energy resolution is necessary to definitively draw this conclusion. We recognize FER only approximates the location of the work function, and more experimental and theoretical investigations are necessary to tease apart the complexity of the underlying physical phenomena at play.

**Conclusions**

In conclusion, this study confirms that environmental proximity engineering is a powerful method to noninvasively create a nanoscale, planar, lateral heterojunction by coupling a TMD monolayer with a nanostructured electrostatic potential template. We show that intercalation can be used as a tool to create heterojunctions by utilizing a facile approach of all UHV *in situ* epitaxial growth and intercalation with Se vapor. We further mapped the band profile, band gap renormalization, and work function of the MoSe$_2$ heterojunction at the nanoscale using STM/S, giving a complete spatial characterization of their relationships. This characterization reveals that the heterojunction has a built-in potential of nearly 1 eV at the nanometer length scale, larger than previously reported TMD lateral heterojunctions,[3, 20, 21] which suggests the possibility of devices that can function at room temperature. In addition to enabling the creation of an ultra-narrow junction, this approach further minimizes the effect of dopant scattering. This demonstration of a nanoscale heterojunction is surprising in light of a previous theoretical analysis that predicts that the depletion region for a doped p-n junction is significantly wider for 2D heterojunctions than that of 3D counterparts.[38] Thus, the creation of a nanoscale heterojunction with a built-in potential of ~0.83 eV exemplifies that proximity engineering can expand the limits of attainable band profiles for TMD based devices.  In fact, work function shifts greater than 2 eV have been demonstrated for other hBN/intercalant/substrate systems,[39] indicating that even larger built-in potentials are most likely achievable.  We expect this approach is not limited to the scope of hBN/TM templates and can be generalized to any substrate or capping layer with nanostructured electrostatic properties, inviting further exploration for nanoscale devices based on atomically thin semiconductors.

**Methods**

**Growth and Intercalation of MoSe$_2$/hBN/Ru Heterostructures.** A clean Ru(0001) surface was prepared by sputtering a Ru single crystal with Ar$^+$ and annealing at 1000 °C in repeated cycles (1.5 kV, 5 × 10$^{-5}$ Torr Ar). A monolayer of hBN was epitaxially grown on the Ru by leaking a borazine (H$_6$B$_3$N$_3$) precursor vapor into a UHV chamber at a pressure of 1 × 10$^{-7}$ Torr for 5 minutes while the Ru was held at 1000 °C, then slowly lowered to room temperature. The quality of the hBN was checked by *in situ* STM and LEED measurements. The hBN/Ru was transferred *in situ* to an MBE chamber for the intercalation and MoSe$_2$ growth. To prepare the intercalated hBN/Ru sample, an effusion cell with Se was used to deposit Se at a rate of 6 Å/min on the hBN surface, at a pressure of 1 × 10$^{-9}$ Torr (the chamber base pressure is 3 × 10$^{-10}$ Torr) for 60 minutes. The sample was maintained at a temperature of 480 °C during the deposition, followed by a 30 minute anneal at the same temperature. The MoSe$_2$ growth was performed in the same chamber with an e-beam evaporation of high-purity Mo (99.95%) and effusion of Se (99.999%) sources at a ratio of 1:30. A hBN/Ru sample maintained at 410 °C for a growth of 55 minutes, then the temperature of the Mo source was decreased while the evaporation of the Se continued. The sample was annealed in the Se vapor at the growth temperature for an additional 15 minutes. We note that the intercalated MoSe$_2$/hBN/Ru sample was achieved by growing the MoSe$_2$ on a hBN/Ru substrate that was not Se intercalated beforehand; the intercalation occurred during the growth and anneal of the MoSe$_2$.

**Scanning Tunneling Microscopy and Spectroscopy.** All STM measurements reported here were acquired at a temperature of 77 °K in UHV (chamber base pressure was below 5 × 10$^{-11}$ Torr). Electrochemically etched tungsten tips were cleaned by a UHV electron bombardment treatment. A homebuilt STM was used, and the bias was applied to the sample. The dI/dV spectra were taken with a constant tip to sample distance while the feedback was turned off. The dZ/dV spectra were acquired with the feedback on and the tip sample distance Z changing to maintain a constant current during the bias sweep.

ASSOCIATED CONTENT

**Supporting Information Available:** First principles calculation of the intercalated hBN/Se/Ru, determining the work function with field emission resonance spectroscopy.

AUTHOR INFORMATION

**Corresponding Author**

Chih-Kang Shih − Department of Physics, The University of Texas at Austin, Austin, Texas 78712, United States; orcid.org/0000-0003-2734-7023; Email: shih@ physics.utexas.edu

**Author Contributions**

C.-K.S conceived the idea and designed the experiment. The experiments were carried out initially by Y.C. and C.Z. and completed by M.A.H., H.K., L.F. and M.L.  DFT calculations were carried out by C.-R.P. and M.-Y.C. M.A.H. and C.-K.S. analyzed the experimental data and wrote the manuscript with inputs from all co-authors.

ACKNOWLEDGMENT

This work was primarily supported by NSF MRSEC under Cooperative Agreement No. DMR-1720595.  Additional financial supports from NSF DMR-1808751, Welch Foundation F-1672, and the U.S. Air Force FA2386-18-1-4097 are also acknowledged.  MYC and CRP acknowledges support from the Academia Sinica Postdoctoral Scholar program.

REFERENCES

1.    Jariwala, D.; Sangwan, V. K.; Louhon, L. J.; Marks, T. J.; Hersam, M. C., Emerging Device Applications for Semiconducting Two-Dimensional Transition Metal Dichalcogenides. *ACS Nano* **2014,** *8*, 19.

2.    Zhang, C.; Li, M. Y.; Tersoff, J.; Han, Y.; Su, Y.; Li, L. J.; Muller, D. A.; Shih, C. K., Strain Distributions and Their Influence on Electronic Structures of WSe2-MoS2 Laterally Strained Heterojunctions. *Nat Nanotechnol* **2018,** *13* (2), 152-158.

3.    Chu, Y.-H.; Wang, L.-H.; Lee, S.-Y.; Chen, H.-J.; Yang, P.-Y.; Butler, C. J.; Lu, L.-S.; Yeh, H.; Chang, W.-H.; Lin, M.-T., Atomic Scale Depletion Region at One Dimensional MoSe2-WSe2 Heterointerface. *Applied Physics Letters* **2018,** *113* (24).

4.    Herbig, C.; Zhang, C.; Mujid, F.; Xie, S.; Pedramrazi, Z.; Park, J.; Crommie, M. F., Local Electronic Properties of Coherent Single-Layer $WS_2/WSe_2$ Lateral Heterostructures. *Nano Lett* **2021**, *21*.

5.    Pielić, B.; Novko, D.; Šrut Rakić, I.; Cai, J.; Petrović, M.; Ohmann, R.; Vujičić, N.; Basletić, M.; Busse, C.; Kralj, M., Electronic Structure of Quasi-Freestanding $WS_2/MoS_2$ Heterostructures. *ACS Appl Mater Interfaces* **2021**, *13*.

6.    Gong, C.; Zhang, H.; Wang, W.; Colombo, L.; Wallace, R. M.; Cho, K., Band Alignment of Two-Dimensional Transition Metal Dichalcogenides: Application in Tunnel Field Effect Transistors. *Applied Physics Letters* **2013,** *103*.

7.    Lee, S. Y.; Kim, U. J.; Chung, J.; Nam, H.; Jeong, H. Y.; Han, G. H.; Kim, H.; Oh, H. M.; Lee, H.; Kim, H.; Roh, Y. G.; Kim, J.; Hwang, S. W.; Park, Y.; Lee, Y. H., Large Work Function Modulation of Monolayer MoS2 by Ambient Gases. *ACS Nano* **2016,** *10* (6), 6100-7.

8.    Allain, A.; Kang, J.; Banerjee, K.; Kis, A., Electrical Contacts to Two-Dimensional Semiconductors. *Nat Mater* **2015,** *14* (12), 1195-205.

9.    Dendzik, M.; Bruix, A.; Michiardi, M.; Ngankeu, A. S.; Bianchi, M.; Miwa, J. A.; Hammer, B.; Hofmann, P.; Sanders, C. E., Substrate-Induced Semiconductor To-Metal Transition in Monolayer WS2. *Physical Review B* **2017,** *96* (23).

10.   Bruix, A.; Miwa, J. A.; Hauptmann, N.; Wegner, D.; Ulstrup, S.; Grønborg, S. S.; Sanders, C. E.; Dendzik, M.; Grubišić Čabo, A.; Bianchi, M.; Lauritsen, J. V.; Khajetoorians, A. A.; Hammer, B.; Hofmann, P., Single-layerMoS2on Au(111): Band Gap Renormalization and Substrate Interaction. *Physical Review B* **2016,** *93* (16).

11.   Zhang, Q.; Chen, Y.; Zhang, C.; Pan, C. R.; Chou, M. Y.; Zeng, C.; Shih, C. K., Bandgap Renormalization and Work Function Tuning in MoSe2/hBN/Ru(0001) Heterostructures. *Nat Commun* **2016,** *7*, 13843.

12.   Ulstrup, S.; Giusca, C. E.; Miwa, J. A.; Sanders, C. E.; Browning, A.; Dudin, P.; Cacho, C.; Kazakova, O.; Gaskill, D. K.; Myers-Ward, R. L.; Zhang, T.; Terrones, M.; Hofmann, P., Nanoscale Mapping of Quasiparticle Band Alignment. *Nat Commun* **2019,** *10* (1), 3283.

13.   Le Quang, T.; Cherkez, V.; Nogajewski, K.; Potemski, M.; Dau, M. T.; Jamet, M.; Mallet, P.; Veuillen, J. Y., Scanning Tunneling Spectroscopy of Van Der Waals Graphene/Semiconductor Interfaces: Absence of Fermi Level Pinning. *2D Materials* **2017,** *4* (3).

14.   Ugeda, M. M.; Bradley, A. J.; Shi, S. F.; da Jornada, F. H.; Zhang, Y.; Qiu, D. Y.; Ruan, W.; Mo, S. K.; Hussain, Z.; Shen, Z. X.; Wang, F.; Louie, S. G.; Crommie, M. F., Giant Bandgap Renormalization and Excitonic Effects in a Monolayer Transition Metal Dichalcogenide Semiconductor. *Nat Mater* **2014,** *13* (12), 1091-5.

15.    Komsa, H.-P.; Krasheninnikov, A. V., Effects of Confinement and Environment on the Electronic Structure and Exciton Binding Energy of MoS2 From First Principles. *Physical Review B* **2012,** *86* (24).

16.    Raja, A.; Waldecker, L.; Zipfel, J.; Cho, Y.; Brem, S.; Ziegler, J. D.; Kulig, M.; Taniguchi, T.; Watanabe, K.; Malic, E.; Heinz, T. F.; Berkelbach, T. C.; Chernikov, A., Dielectric Disorder in Two-Dimensional Materials. *Nat Nanotechnol* **2019,** *14* (9), 832-837.

17.    Stier, A. V.; Wilson, N. P.; Clark, G.; Xu, X.; Crooker, S. A., Probing the Influence of Dielectric Environment on Excitons in Monolayer WSe2: Insight from High Magnetic Fields. *Nano Lett* **2016,** *16* (11), 7054-7060.

18.    Waldecker, L.; Raja, A.; Rosner, M.; Steinke, C.; Bostwick, A.; Koch, R. J.; Jozwiak, C.; Taniguchi, T.; Watanabe, K.; Rotenberg, E.; Wehling, T. O.; Heinz, T. F., Rigid Band Shifts in Two-Dimensional Semiconductors through External Dielectric Screening. *Phys Rev Lett* **2019,** *123* (20), 206403.

19.    Raja, A.; Chaves, A.; Yu, J.; Arefe, G.; Hill, H. M.; Rigosi, A. F.; Berkelbach, T. C.; Nagler, P.; Schuller, C.; Korn, T.; Nuckolls, C.; Hone, J.; Brus, L. E.; Heinz, T. F.; Reichman, D. R.; Chernikov, A., Coulomb Engineering of the Bandgap and Excitons in Two-Dimensional Materials. *Nat Commun* **2017,** *8*, 15251.

20.    Utama, M. I. B.; Kleemann, H.; Zhao, W. Y.; Ong, C. S.; da Jornada, F. H.; Qiu, D. Y.; Cai, H.; Li, H.; Kou, R.; Zhao, S. H.; Wang, S.; Watanabe, K.; Taniguchi, T.; Tongay, S.; Zettl, A.; Louie, S. G.; Wang, F., A Dielectric-Defined Lateral Heterojunction in a Monolayer Semiconductor. *Nat. Electron.* **2019,** *2* (2), 60-65.

21.    Song, Z.; Schultz, T.; Ding, Z.; Lei, B.; Han, C.; Amsalem, P.; Lin, T.; Chi, D.; Wong, S. L.; Zheng, Y. J.; Li, M. Y.; Li, L. J.; Chen, W.; Koch, N.; Huang, Y. L.; Wee, A. T. S.,

Electronic Properties of a 1D Intrinsic/p-Doped Heterojunction in a 2D Transition Metal Dichalcogenide Semiconductor. *ACS Nano* **2017,** *11* (9), 9128-9135.

22.    Chen, J. W.; Lo, S. T.; Ho, S. C.; Wong, S. S.; Vu, T. H.; Zhang, X. Q.; Liu, Y. D.; Chiou, Y. Y.; Chen, Y. X.; Yang, J. C.; Chen, Y. C.; Chu, Y. H.; Lee, Y. H.; Chung, C. J.; Chen, T. M.; Chen, C. H.; Wu, C. L., A Gate-Free Monolayer WSe2 PN Diode. *Nat Commun* **2018,** *9* (1), 3143.

23.    Laskowski, R.; Blaha, P., Ab Initio Study of h−BN Nanomeshes on Ru(001), Rh(111), and Pt(111). *Physical Review B* **2010,** *81* (7).

24.    Goriachko, A.; He, Y.; Knap, M.; Over, H., Self-Assembly of a Hexagonal Boron Nitride Nanomesh on Ru(0001). *Langmuir* **2007,** *23*, 2928-2931.

25.    Brugger, T.; Ma, H.; Iannuzzi, M.; Berner, S.; Winkler, A.; Hutter, J.; Osterwalder, J.; Greber, T., Nanotexture Switching of Single-Layer Hexagonal Boron Nitride on Rhodium by Intercalation of Hydrogen Atoms. *Angew Chem Int Ed Engl* **2010,** *49* (35), 6120-4.

26.    Dong, A.; Fu, Q.; Wu, H.; Wei, M.; Bao, X., Factors Controlling the CO Intercalation of h-BN Overlayers on Ru(0001). *Phys Chem Chem Phys* **2016,** *18* (35), 24278-84.

27.    Ng, M. L.; Shavorskiy, A.; Rameshan, C.; Mikkelsen, A.; Lundgren, E.; Preobrajenski, A.; Bluhm, H., Reversible Modification of the Structural and Electronic Properties of a Boron Nitride Monolayer by CO Intercalation. *Chemphyschem* **2015,** *16* (5), 923-7.

28.    Wei, M.; Fu, Q.; Wu, H.; Dong, A.; Bao, X., Hydrogen Intercalation of Graphene and Boron Nitride Monolayers Grown on Pt(111). *Topics in Catalysis* **2015,** *59* (5-7), 543-549.

29.    Yang, Y.; Fu, Q.; Wei, M.; Bluhm, H.; Bao, X., Stability of BN/Metal Interfaces in Gaseous Atmosphere. *Nano Research* **2014,** *8* (1), 227-237.

30.    Corso, M.; Auwarter, W.; Muntwiler, M.; Tamai, A.; Greber, T.; Osterwalder, J., Boron Nitride Nanomesh. *Science* **2004,** *303*, 4.

31.    Stolbov, S., First-Principles Study of Formation of Se Submonolayer Structures on Ru Surfaces. *Phys Rev B* **2010**, *82*, 155463.

32.    Stolbov, S., Nature of Selenium Submonolayer Effect on the Oxygen Electroreduction Reaction Activity of Ru(0001). *J Phys Chem C* **2012**, *116*, 7173-7179.

33.    Stroscio, J. A.; Feenstra, R. M.; Fein, A. P., Electronic Structure of the Si(111)2 x 1 Surface by Scanning-Tunneling Microscopy. *Phys Rev Lett* **1986,** *57* (20), 2579-2582.

34.    Feenstra, R. M.; Stroscio, J. A., Tunneling Spectroscopy of the GaAs(110) Surface. *J. Vac.Sci. Technol. B* **1987,** *5* (4), 6.

35.    Zhang, C.; Chen, Y.; Johnson, A.; Li, M. Y.; Li, L. J.; Mende, P. C.; Feenstra, R. M.; Shih, C. K., Probing Critical Point Energies of Transition Metal Dichalcogenides: Surprising Indirect Gap of Single Layer WSe2. *Nano Lett* **2015,** *15* (10), 6494-500.

36.    Steinke, C.; Wehling, T. O.; Rösner, M., Coulomb-Engineered Geterojunctions and Dynamical Screening in Transition Metal Dichalcogenide Monolayers. *Physical Review B* **2020,** *102* (11).

37.    Rosner, M.; Steinke, C.; Lorke, M.; Gies, C.; Jahnke, F.; Wehling, T. O., Two-Dimensional Heterojunctions from Nonlocal Manipulations of the Interactions. *Nano Lett* **2016,** *16* (4), 2322-7.

38.    Ilatikhameneh, H.; Ameen, T.; Chen, F.; Sahasrabudhe, H.; Klimeck, G.; Rahman, R., Dramatic Impact of Dimensionality on the Electrostatics of P-N Junctions and Its Sensing and Switching Applications. *IEEE Transactions on Nanotechnology* **2018,** *17* (2), 293-298.

39.    Fedorov, A.; Praveen, C. S.; Verbitskiy, N. I.; Haberer, D.; Usachov, D.; Vyalikh, D. V.; Nefedov, A.; Wöll, C.; Petaccia, L.; Piccinin, S.; Sachdev, H.; Knupfer, M.; Büchner, B.; Fabris, S.; Grüneis, A., Efficient Gating of Epitaxial Boron Nitride Monolayers by Substrate Functionalization. *Physical Review B* **2015,** *92* (12).

TOC Image

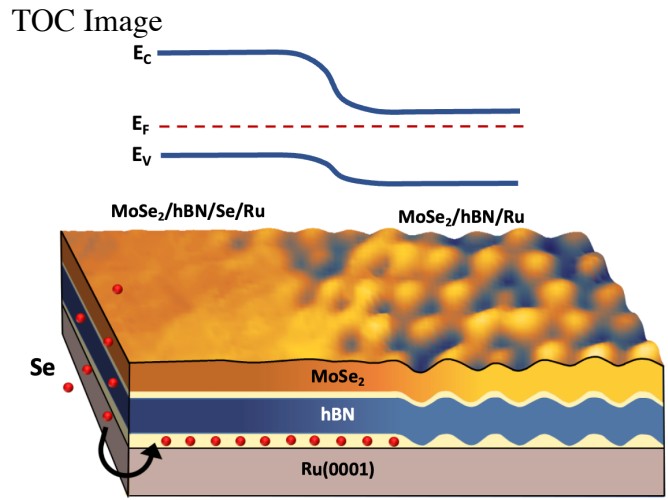